# Compact USB measurement and analysis system for real-time fluctuation enhanced sensing

Robert Mingesz, Zoltán Gingl
Department of Technical Informatics
University of Szeged
Szeged, Hungary
rmingesz@titan.physx.u-szeged.hu

Ákos Kukovecz, Zoltán Kónya
Department of Applied and Environmental Chemistry
University of Szeged
Szeged, Hungary

Krisztián Kordás, Hannu Moilanen
Microelectronics and Materials Physics Laboratories
University of Oulu
Oulu, Finland

*Abstract*—Measuring the resistance fluctuations of gas sensors provides new opportunities to enhance the selectivity and sensitivity of the sensor. Taking advantage of this possibility requires special low-noise measurement hardware and software to acquire data and perform analysis. In our talk we will present a small, USB-powered device capable of doing precise measurement of the resistance fluctuations of different kinds of gas sensors. We have developed a graphical user interface software to control the parameters of the measurement, to collect data and perform real time analysis on the measured data. The analysis is based on a PCA algorithm, which is proven to be a high performance tool to support fluctuation enhanced sensing. The system has been tested on Taguchi and carbon nanotube based gas sensors as well. The main advantages of the system include the small form factor, low cost and the fully featured software performing all required data analysis operations. Complemented with a gas sensor and an optional test chamber, the setup can serve as an efficient tool for practical fluctuation-enhanced gas sensing.

## I. INTRODUCTION

Fluctuation enhanced sensing is an emerging way of getting out more information of the output signal of some kinds of sensors with the help of intensive analysis of the random fluctuations [1-3]. It is well known that noise coming from a system can carry important information about the state and behavior of the subject;, however, it is not always straightforward to find the experimental and signal processing tools for extracting this information. Safety, home automation, industrial analysis and control are some reasons of growing gas sensor applications. Commercially available Taguchi gas sensors and promising nanotechnology-based gas sensors show change in their resistance as a function of the concentration of different externally applied gases. Measuring only the average of this resistance gives only very limited information, especially in the presence of a mixture of several gases, while it has been shown that noise analysis can give significantly more information, has the potential to identify the type and concentration of the surrounding gases in some cases.

Developing noise measurement techniques and signal processing methods can help to improve the quality and efficiency of gas sensing, and it is also aimed at making low-power, portable devices and intelligent sensor networks.

In most cases a comprehensive set of professional instruments - like low-noise preamplifiers, dynamic signal analyzers - are used with some proposals of developing compact, dedicated devices to perform the measurement and analysis [4-6]. In the following we briefly present a small, low-power device and the accompanying analysis software developed especially to support fluctuation enhanced gas sensing.

## II. USB-POWERED DATA ACQUISITION UNIT WITH GAS SENSOR INTERFACE

### A. System setup

The system consists of an analog signal processing part and a mixed-signal processor to convert the signals into the digital domain, and communicates with the host computer via a universal asynchronous receiver-transmitter (UART)-to-USB interface.

The block diagram of the system is shown in Fig. 1.

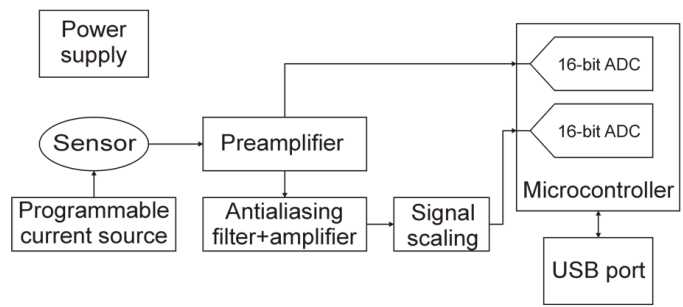

Figure 1. Block diagram of the system.

This work was supported by grant OTKA K69018 of the Hungarian Academy of Sciences

## B. Gas sensor interface and analog signal processing

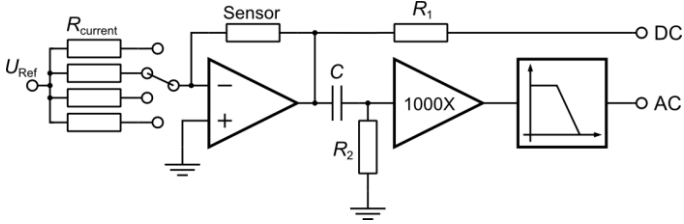

Figure 2. Simplified schematic of the gas sensor interface. The DC value is proportional to the sensor's resistance while the AC value represents the fluctuation of the sensor resistance.

There resistance and resistance fluctuations of the gas sensor can be converted to voltage by the use of a high precision, low noise current source. The current source uses a precision voltage reference (AD780) and a MAX4478 low voltage and current noise operational amplifier. The operational amplifier is used in an inverting configuration: the properly filtered reference voltage is connected to the inverting input via an analog multiplexer-selectable series resistor, while the sensor is placed in the feedback loop. The output voltage of this amplifier is proportional to the sensor resistance, whose average magnitude and fluctuation depends on the concentration of the applied gas. A two-stage active high-pass filter removes the DC component and amplifies the voltage fluctuations by 1000. The last component of the analog signal conditioning chain is a programmable gain amplifier and high-precision anti-aliasing filter (LTC1564).

## C. Data acquisition and USB interface

The main component of the data acquisition system is a precision mixed-signal microcontroller (C8051F060). The two output voltages of the analog signal processing part is digitized simultaneously by the microcontroller's dual 16-bit analog-to-digital converters. The on-chip SRAM has been used as a temporary buffer for digitized data to ensure continuous, real-time sampling and data transfer to the host computer via the UART-USB interace chip (FT232RL). The sampling frequency can be set up to 50 kHz ensuring maximum measured signal bandwith of 20 kHz, adequate for fluctuation-enhanced sensing applications.

The supply options include powering the device from the USB port or an external DC supply by the use of a low-noise, properly filtered DC/DC converter (TMR2-0521).

Gas sensors often need heating voltage up to 5 V; in our system, the microcontroller's built-in 12-bit digital-to-analog converter set the heating voltage to the desired value. A second digital-to-analog converter can be used to set the optional gate voltage of the sensor.

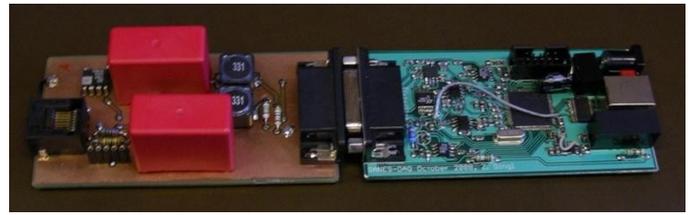

Figure 3. The analog signal processing unit (left) and the USB data acquisition system connected together.

## D. Microcontroller software

The microcontroller runs a simple software that configures the analog preprocessing circuitry, sets the sample rate and controls the analog-to-digital conversion. The host communication is made easy by interpreting bytes received from the host as commands. The microcontroller does not make any digital signal processing, only transfers the raw digitized data to the host computer for analysis.

## III. ANALYSIS SOFTWARE

The graphical user interface software runs on the host computer; its primary aim is to provide a complete fluctuation enhanced sensing solution. The software communicates with the microcontroller of the data acquisition system via the USB port. It enables full control of the system and measurement parameters including the sample rate, anti-aliasing filter cutoff frequency, sensor excitation current, heating voltage and sensor gate voltage. The software can also receive the continuous flow of measurement data in real time for on-line processing.

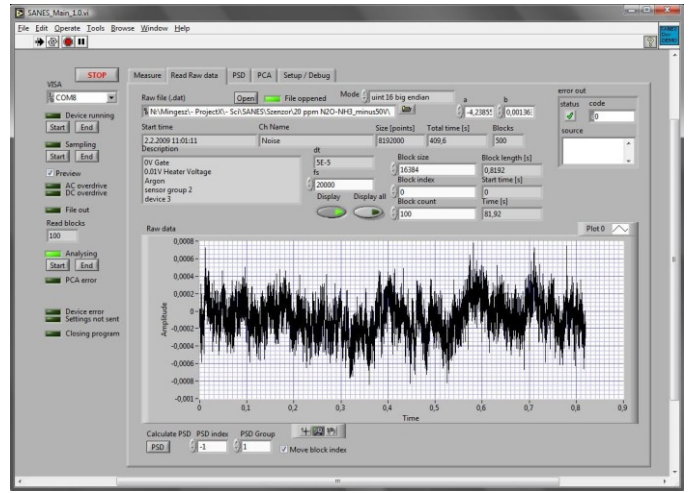

Figure 4. Screenshot of the software showing raw measured data display. There are many controls and indicators for setting up the measurement at the highest possible flexibility.

The analysis of the data is based on calculation of the power spectral density (PSD). The user can define an unlimited number of averages, while the frequency range is set by selecting the values for the sample rate and the total time of the sequence of samples. There is an option to remove the fundamental and the harmonics of the 50 or 60 Hz from the

PSD. Since the PSD can be considered as the raw input for any other signal processing, the measured PSDs can be saved into a file and can be loaded later to perform comparison or custom analysis.

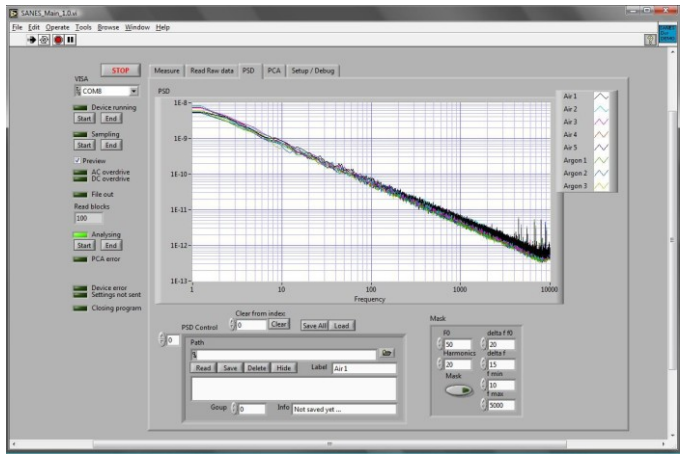

Figure 5. Display of the measured and stored PSDs.

The main advantage of the fluctuation enhanced sensing method is the improved selectivity and sensitivity to different gases even using a single sensor; however, extensive signal processing is needed. There are many possibilities of trying to extract the information from the measured PSD data, in our implementation we applied principal component analysis (PCA). This method tries to find an optimal projection of a set of PSDs in which the obtained values corresponding to individual PSDs are the most distinguishable in a two dimensional space. The PCA calculation can be performed on-line, while still averaging the PSD-s.

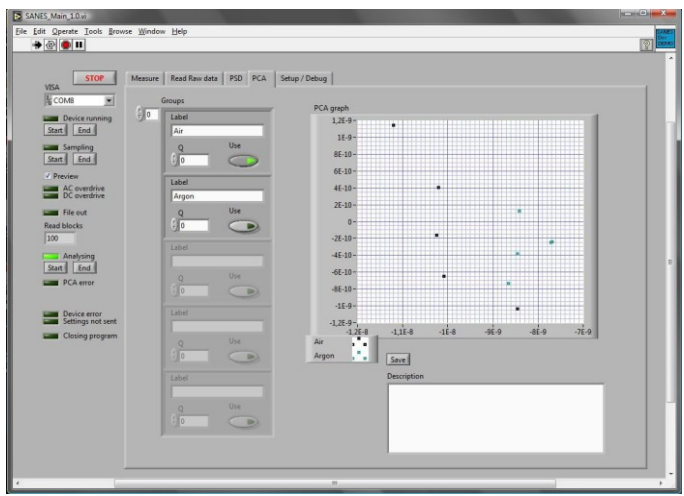

Figure 6. PCA analysis tab. The plot shows the result of PCA calculations. Each point corresponds to a measured PSD. As more and more PSDs have been acquired with different gases, a basis can be built up to identify an unknown condition from the measured PSD.

## IV. EXAMPLE EXPERIMENTAL RESULTS

The developed measurement hardware and software has been tested on carbon nanotube (CNT) sensors. The thin film sensor resistance varied between 10 kOhms and 1 MOhms, both the resistance and its fluctuations were measured. Different gases and concentrations were applied including CO, NO, $CH_4$ in argon buffer in a chamber providing precise control of gas concentration and gas flow.

Figure 7 shows the result of PCA analysis for different types of gases applied separately. The sample rate was 20 kHz, 16384 points were used to calculate the PSDs and 100 PSDs were averaged prior performing the PCA analysis. Five consecutive measurements were carried out for each gas type. It can be seen that the result of the PCA analysis allows separation of the different gases using a single sensor.

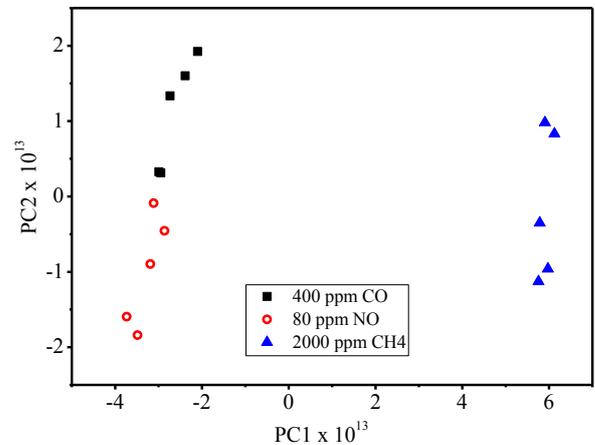

Figure 7. PCA analysis for different type gases using a single CNT sensor. The sensor is functionalized with carboxyl, the temperature was 25°C, the buffer gas was argon.

We also demonstrate system performance using a single CNT sensor and CO gas with different concentration values. The same measurement parameters were used as shown above; after measuring five PSDs the PCA analysis were applied. The result can be seen of Figure 7, illustrating the selectivity of the system.

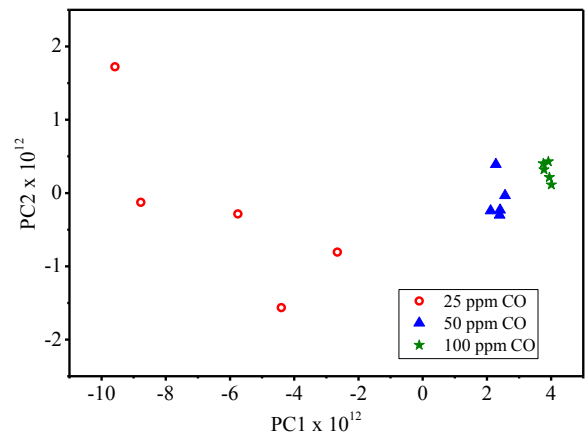

Figure 8. PCA analysis for different concentration of CO gas using a single sulfonated CNT sensor. The temperature was 80°C, the buffer gas was argon.

## V. CONCLUSION

A small, compact USB powered device has been developed for fluctuation enhanced gas sensing applications. The system is complete, it includes the gas sensor excitation low noise analog signal conditioning electronics, anti-aliasing filters and a mixed signal microcontroller with built-in 16-bit analog-to-digital converters. The measurement system is connected to the host computer via the USB interface. A graphical user interface software application controls the whole measurement process and allows real-time noise measurement and spectral analysis. A PCA algorithm has been implemented to extract information about the type and concentration of the applied gas. The whole system has been tested on CNT gas sensors with different gases and concentrations.

## ACKNOWLEDGMENT

The authors thank Laszlo B. Kish and Peter Makra for useful discussions.